\documentclass[journal]{IEEEtran}

\IEEEoverridecommandlockouts

\usepackage[left=2cm,right=2cm]{geometry}
\usepackage[utf8]{inputenc}
\usepackage{xcolor}
\usepackage{hyperref}
\usepackage{xspace}
\usepackage{color}
\usepackage{listings}
\usepackage{acronym}
\usepackage[T1]{fontenc}
\usepackage{lmodern}
\usepackage{multirow}
\usepackage{colortbl}
\usepackage{todonotes}
\usepackage{url}
\usepackage[export]{adjustbox}
\usepackage{tcolorbox}
\usepackage{comment}
\usepackage{colortbl}
\usepackage{tcolorbox}
\usepackage{lipsum}
\usepackage{multicol}
\usepackage{amsmath}
\usepackage{amssymb,amsfonts}
\usepackage{float}
\usepackage{graphicx}
\usepackage{algpseudocode}
\usepackage{algorithm}
\usepackage{listings}
\usepackage[flushleft]{threeparttable}
\usepackage{subcaption}%

\definecolor{codegreen}{rgb}{0,0.6,0}
\definecolor{codegray}{rgb}{0.5,0.5,0.5}
\definecolor{codepurple}{rgb}{0.58,0,0.82}
\definecolor{backcolour}{rgb}{0.95,0.95,0.92}
\definecolor{lightgray}{rgb}{.9,.9,.9}
\definecolor{darkgray}{rgb}{.4,.4,.4}
\definecolor{darkgreen}{rgb}{0, 0.39, 0.00}
\definecolor{Gray}{gray}{0.7}

\lstdefinestyle{mystyle}{
    backgroundcolor=\color{backcolour},   
    commentstyle=\color{codegreen},
    keywordstyle=\color{magenta},
    numberstyle=\tiny\color{codegray},
    stringstyle=\color{codepurple},
    basicstyle=\ttfamily\footnotesize,
    breakatwhitespace=false,         
    breaklines=true,                 
    captionpos=b,                    
    keepspaces=true,                 
    numbers=left,                    
    numbersep=5pt,                  
    showspaces=false,                
    showstringspaces=false,
    showtabs=false,                  
    tabsize=2
}

\acrodef{ECU}{Electronic Control Unit}
\acrodef{CACC}{Cooperative Adaptive Cruise Control}
\acrodef{RSU}{Road Side Unit}
\acrodef{SC}{Service Center}
\acrodef{ITS}{Intelligent Transportation System}
\acrodef{OBU}{On-Board Unit}
\acrodef{OBD}{On-Board Diagnostics}
\acrodef{CAN}{Controller Area Network}
\acrodef{DoS}{Denial of Service}
\acrodef{AAA}{Automobile Association of America}
\acrodef{V2V}{Vehicle-to-Vehicle}
\acrodef{V2I}{Vehicle-to-Infrastructure}
\acrodef{PID}{Priority ID}
\acrodef{CIDS}{Clock-based IDS}
\acrodef{RLS}{Recursive Least Squares}
\acrodef{IDS}{Intrusion Detection System}
\acrodef{IVI}{In Vehicle Infotainment}
\acrodef{HMM}{Hidden Markov Model}
\acrodef{ML}{Machine Learning}
\acrodef{SAE}{Society of Automotive Engineers}
\acrodef{RPM}{Revolutions Per Minute}
\acrodef{CSMA/CD}{Carrier Sense Multiple Access with Collision Detection} 
\acrodef{ML}{Machine Learning}
\acrodef{NN}{Neural Networks}
\acrodef{TPM}{Tire Pressure Monitoring System}
\acrodef{GPS}{Global Positioning System}
\acrodef{DNN}{Deep Neural Network}
\acrodef{LSTM}{Long Short-Term Memory}
\acrodef{MSG}{Messages-Sequence Graph}
\acrodef{CPD}{Change Point Detection}
\acrodef{RNN}{Recurrent Neural Networks}
\acrodef{GID}{Generative Adversarial Nets based Intrusion  Detection System}
\acrodef{CNN}{Convolution Neural Networks}
\acrodef{GRU}{Gated Recurrent Unit}
\acrodef{GAN}{Generative Adversarial Nets}
\acrodef{OBD}{On-Board Diagnostics}
\acrodef{RT}{Real-Time}
\acrodef{ACC}{Adaptive Cruise Control }
\acrodef{DDoS}{Distribute Denial of Service}
\acrodef{CPS}{Cyber Physical System}
\acrodef{IDS}{Intrusion Detection System}
\acrodef{CAN}{Controller Area Network}
\acrodef{HMM}{Hodden Markov Model}
\acrodef{SAE}{Society of Automotive Engineers}
\acrodef{IPC}{Inter-Process Communication}

\title{Evaluation of the Architecture Alternatives for Real-time Intrusion Detection Systems for Connected Vehicles}
\author{\IEEEauthorblockN{Mubark B Jedh}

\and
\IEEEauthorblockN{Jian Kai Lee}

\and
\IEEEauthorblockN{Lotfi ben Othmane}

}

\author{\IEEEauthorblockN{Mubark Jedh
\IEEEauthorrefmark{1}, Jian Kai Lee\IEEEauthorrefmark{1},  Lotfi ben Othmane\IEEEauthorrefmark{1}}
\\
\IEEEauthorblockA{\IEEEauthorrefmark{1} Iowa State University, Ames, IA USA}
}

\begin{document}

\maketitle

\begin{abstract}

Attackers demonstrated the use of remote access to the in-vehicle network of connected vehicles to launch cyber-attacks and remotely take control of these vehicles. Machine-learning-based \acp{IDS} techniques have been proposed for the detection of such attacks. The evaluation of some of these \ac{IDS} demonstrated their efficacy in terms of accuracy in detecting message injections but was performed offline, which limits the confidence in their use for real-time protection scenarios. This paper evaluates four architecture designs for real-time \ac{IDS} for connected vehicles using \ac{CAN} datasets collected from a moving vehicle under malicious speed reading message injections. The evaluation shows that a real-time \ac{IDS} for a connected vehicle designed as two processes, a process for CAN Bus monitoring and another one for anomaly detection engine is reliable (no loss of messages) and could be used for real-time resilience mechanisms as a response to cyber-attacks.

\end{abstract}

\section{Introduction}

\begin{figure}[tbp]
\centering
	\includegraphics[width=0.45\textwidth]{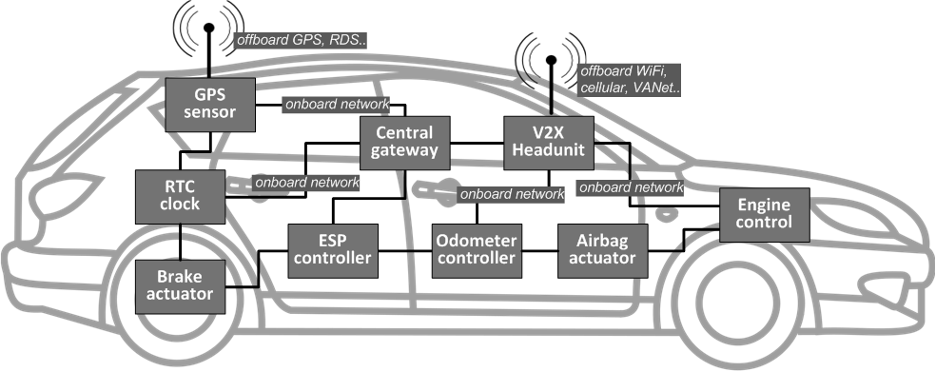} 
\caption{Example of connected vehicle~\cite{Othmane2015}}
\label{fig:connectvehicle}
\end{figure}

Modern Automobile contains 20 to 80 \acp{ECU} that control the functionalities of the vehicle, including engine, power steering, brake, air-conditioning, radio, and seats, lane departure alert, and emergency braking, as depicted by Figure~\ref{fig:connectvehicle}. These \acp{ECU} communicate using the \acf{CAN} bus protocol~\cite{bosh1991}. The \ac{CAN} protocol was designed more than 40 years ago and has still widespread use in automotive, aerospace, and many other industries because of its low cost, error detection capability, reliability, and high-speed transmission rate. The \ac{CAN} protocol does not include, however, security measures such as authentication~\cite{OWMW2015}.

  \begin{figure*}[tbp]
  \centering
  \begin{subfloat}[Increase of cyber-attacks on connected vehicles.]
        {\includegraphics[width=0.42\linewidth]{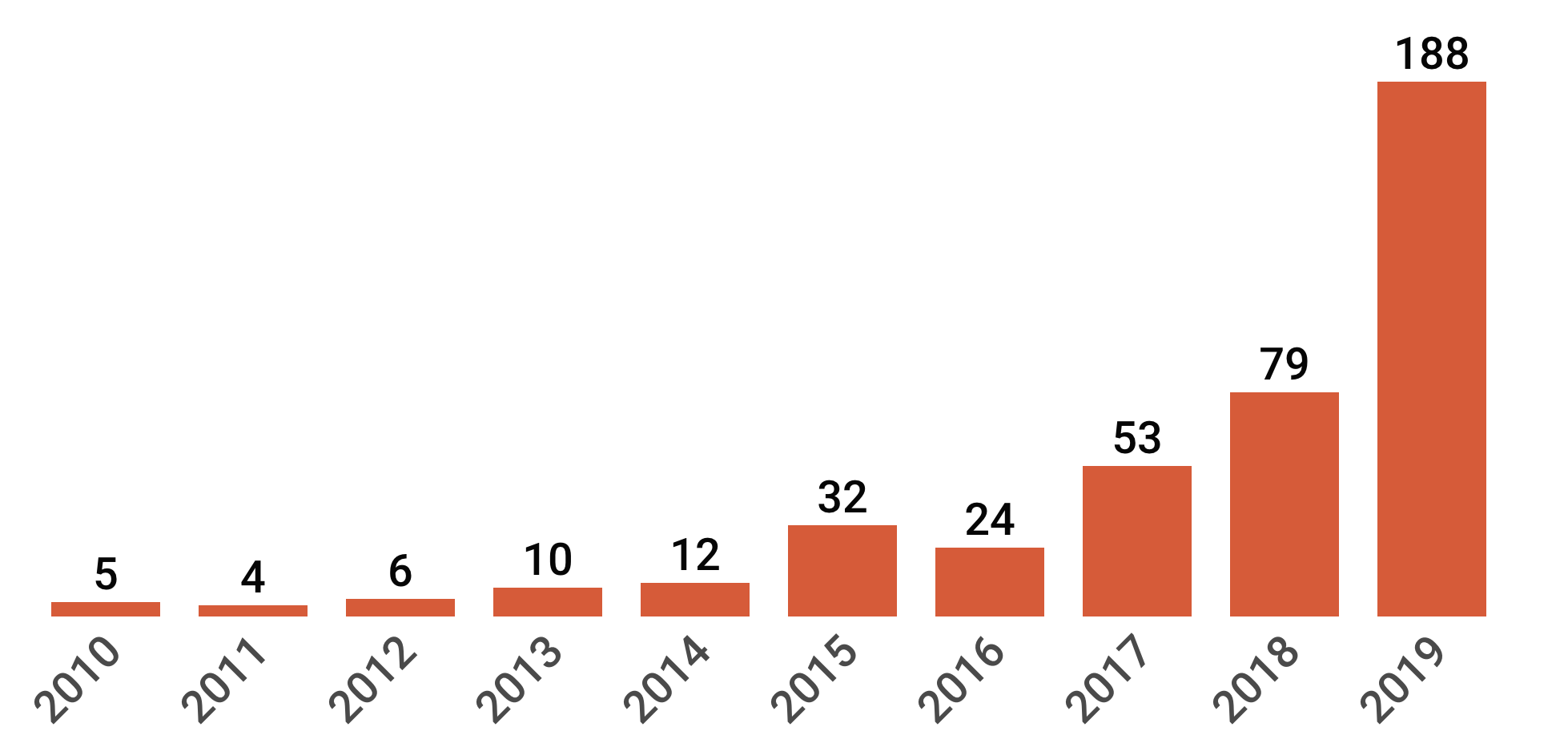}\label{Figures:incidentgrowth}}
       %
  \end{subfloat} 
 ~
    \begin{subfloat}[Impacts of cyber-attacks on connected vehicles]
       {\includegraphics[width=0.55\linewidth]{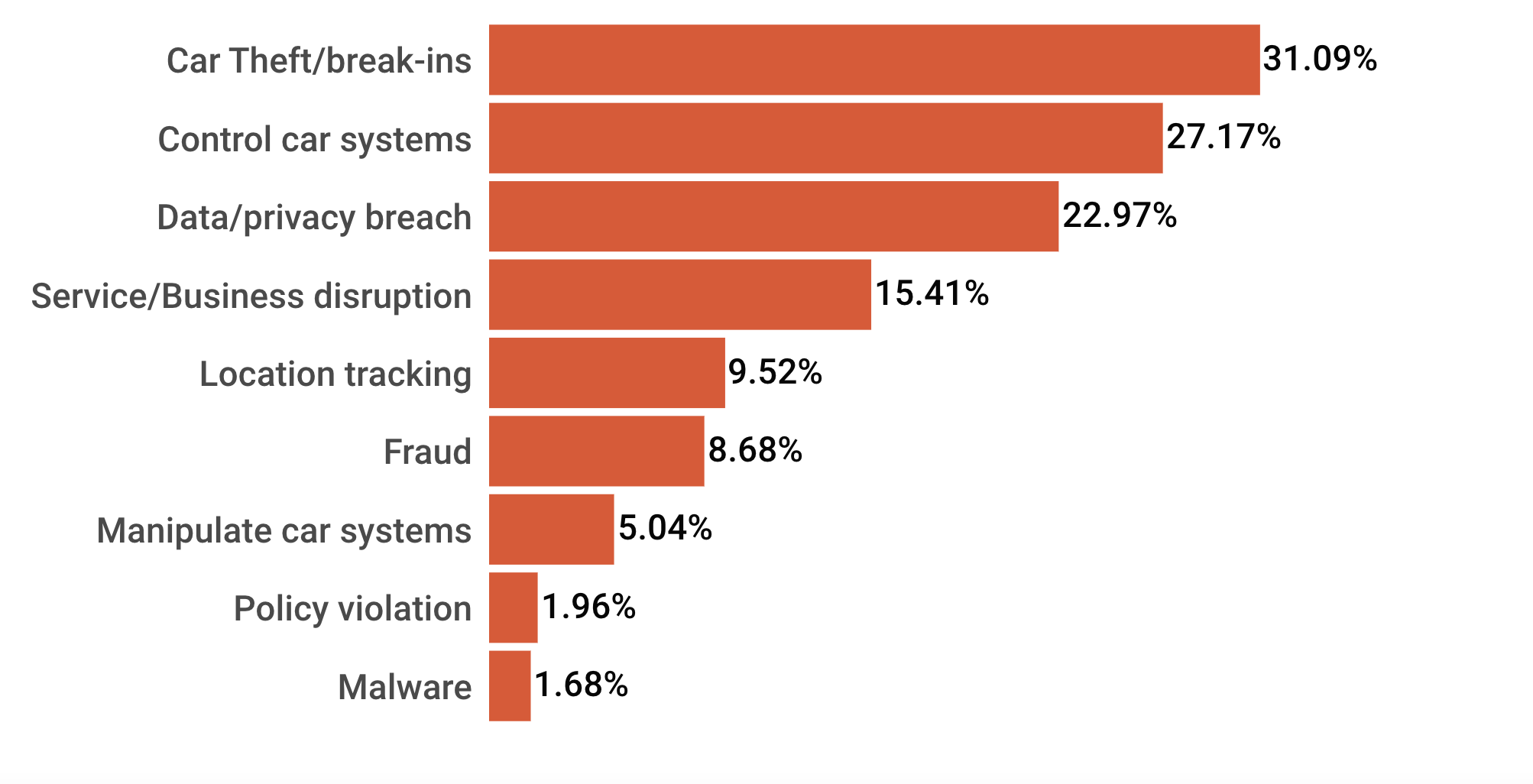}\label{fig:incidentimpact}}

  \end{subfloat}  
  \caption{Growth and distribution of cyber-attacks on connected vehicles between 2010 and 2019~\cite{UpstreamAuto2020}.}
  \end{figure*}

Several recent research works demonstrate that attackers can access the CAN Bus using a variety of interfaces such as TPMS, bluetooth, telematics, and OBD-II units to inject messages into the CAN Bus. Hoppe et al.~\cite{10.1007/978-3-540-87698-4_21} were the first researchers to point out the security weaknesses of the CAN Bus protocol. Their findings were later confirmed by Koscher et al.~\cite{5504804}. Furthermore, Checkoway et al.~\cite{10.5555/2028067.2028073} performed a security analysis of attack surfaces, including physical, short-range and long-range wireless communication, and demonstrated the exploitation of the flaws that they have identified to fully take control of the vehicle's systems. Recently, Upstream's research team identified 367 publicly reported incidents for a decade long~\cite{UpstreamAuto2020}. The analysis of these incidents shows an exponential growth of attacks, as depicted by Figure~\ref{Figures:incidentgrowth}. Among these attacks, 27\% involved taking control of the car, as depicted by Figure~\ref{fig:incidentimpact}. 

\acfp{IDS} have been proposed as an alternative to the attack prevention approach on connected vehicles. Wu et al.~\cite{8688625} and Young et al.~\cite{8640808} provide comprehensive surveys on \ac{IDS} for connected and autonomous cars. In previous work~\cite{9076852, 9490207} we developed \ac{ML}-based \acp{IDS} for connected vehicles and evaluated them using \ac{CAN} data extracted from a moving vehicle under malicious RPM and speed readings messages injections into the in-vehicle network of the vehicles. Most of the ML-based \ac{IDS} for the connected vehicle are evaluated offline using datasets of \ac{CAN} logs, including the ones that we designed, which limits the confidence in the capabilities to use them for real-time intrusion detection in vehicles. 

 \begin{table}[tbp]
\caption{Major requirements for \ac{IDS} for connected vehicles.}
\label{tab:majorRequirements}
\centering
\begin{tabular}{p{0.1in} p{2.9in}}
\rowcolor{Gray}\hline
  ID & Requirements\\
\hline
 1 & The response time of \ac{IDS} must be small enough to trigger reactive safety mechanisms, such as breaking. \\\hline
 2 & The \ac{IDS} must not lose \ac{CAN} data. \\\hline
 3 & The \ac{IDS} must run on an \acp{ECU} that has limited capabilities in terms of processing speed and memory size.\\\hline
\hline
\end{tabular}
\end{table}

There are three requirements for \acp{IDS} for connected vehicles, which are listed in Table~\ref{tab:majorRequirements}. We describe in this paper four architecture alternatives for real-time \ac{IDS} for connected vehicles.\footnote{We focus on CAN message injection attacks. Other attacks on connected vehicles, e.g., on V2V could be considered in the future.} Then, we assess their satisfaction of the requirements by evaluating their (1) anomaly evaluation time, and (2) reliability in terms of losing CAN messages. The findings demonstrate the possibility to deploy effective ML-based \acp{IDS} for connected vehicles.

This paper is organized as follows. Section~\ref{sec:relworks} describes the related works, section~\ref{sec:IDS} describes the used \ac{IDS}, Section~\ref{sec:arechitecture} describes the architecture alternatives of the real-time \ac{IDS} for connected vehicle, Section~\ref{sec:evaluation} describes the evaluation of the architecture alternatives, and Section~\ref{sec:conclusion} concludes the paper.

\section{Related work}\label{sec:relworks}

Several preventive security countermeasures have been developed to defend and enhance in-vehicle network security against cyber-attacks, such as authentication protocols~\cite{8590911, 8939382, 7934878}. The main issue with these mechanisms is that these address only a subset of the attacks on the connected vehicles and require modification of the protocol used by the \acp{ECU} of the given vehicle, which cannot be used for aftermarket vehicles. In addition, most of the remote attacks exploit software vulnerabilities in the protection mechanisms, such as in~\cite{MiVa2015,hacking-tesla,BMW,Tesla_key}. 
 
\acp{IDS} have been proposed as an alternative to prevention mechanisms from attacks on connected vehicles. Wu et al.~\cite{8688625} and Young et al.~\cite{8640808} provide surveys on \ac{IDS} for connected and autonomous cars. These mechanisms discriminate groups of messages associated with attacks and those that are not, with acceptable accuracy and false positive~\cite{10.1371/journal.pone.0155781}. \ac{NN} has been the commonly used ML-based approach for designing \acp{IDS} for the CAN bus, e.g.,~\cite{Lokman2019, 8687274, 9262960, SGL2019}. In previous work~\cite{9076852, 9490207} we used machine learning techniques to develop \acp{IDS} for connected vehicles including \ac{HMM}, \ac{LSTM}, cosine graph-similarity, and change-point detection and evaluated them using \ac{CAN} data extracted from a moving vehicle under malicious RPM and speed readings messages injections into the in-vehicle network of the vehicles. The detection accuracy of cosine graph-similarity threshold reaches 97.32\% of accuracy and the detection speed of 2.5 milliseconds. Unlike most ML-based studies in the literature, the technique proposed in~\cite{9490207} neither depend on the make or model of the car nor its proprietary information (i.e., CAN ID).

The common data sources that are used to assess the accuracy of the ML-based IDS solutions include; data from the owner devices~\cite{ChSh2016}, synthetic/artificial data ~\cite{TLJ20168}, simulated data~\cite{LMKA2017}, and data from a stationary/parked vehicle~\cite{SHH2018, 9235336, IRYM2020}. This limits the confidence in the results and threatens its validity~\cite{CROT2016}. To the best of our knowledge, our previous work~\cite{9076852, 9490207} and the works of Stachowski~\cite{SGL2019} are the only studies that used datasets collected from moving vehicles under messages injection attacks. 

\begin{figure*}[tbp]
    \begin{subfloat}
        {\includegraphics[width=.50\linewidth]{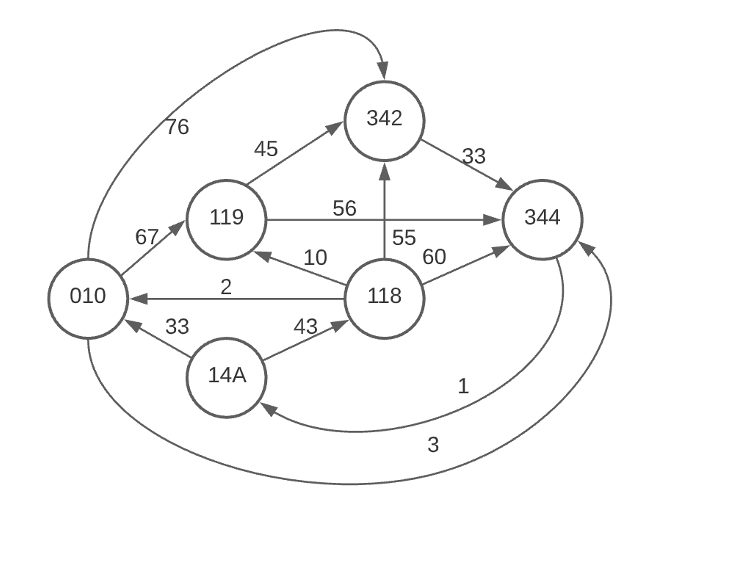}}
    \end{subfloat}
    \begin{subfloat}
        {\includegraphics[width=.50\linewidth]{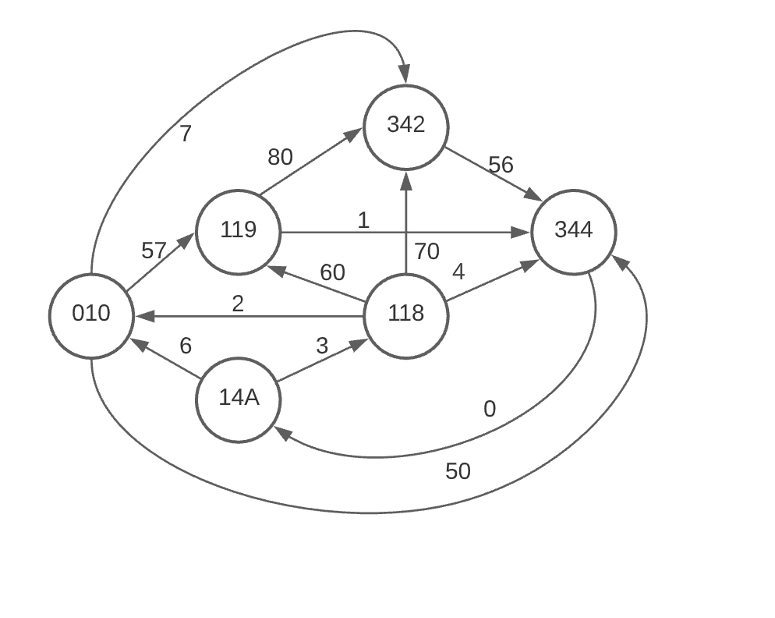}}
    \end{subfloat}
     \caption{Similarity of two messages-sequence graphs at successive time-window $t$ (left) and $t+1$ (right). The labels of the nodes are the CAN ID and the labels of the edges are number of times a message with the CAN ID source of the edge is followed by a message with the CAN ID destination of the same edge during the time-window. For example, 33 messages and 56 messages with ID 344 followed messages with ID 342 at resp. time-windows $t$ and $t+1$, which indicate a possible change of the behavior of the vehicle~\cite{9490207}.
     }
    \label{fig:messagesgraph}
\end{figure*}

Valasek and Miller are among the pioneer to propose real-time \ac{IDS} for connected vehicles~\cite{Chris&Miller}. They developed a small device that reads data from the CAN Bus through OBD-II port, learns the traffic pattern to detect anomalies, and shorts the circuit disabling all CAN messages when anomalous detected. Matsumoto et al.~\cite{6240294} proposed a real-time \ac{IDS} that prevents authorized messages from reaching the receiver ECU. The system monitors the traffic of the CAN Bus and transmits Error Frame to override the unauthorized messages when it detects them. The technique requires, however, modification of the CAN Bus protocol. 
 
Boddupalli and Ray assessed the requirements for real-time attack detection and mitigation in connected and autonomous vehicles and emphasized the importance of the basic safety of such a mechanism. \cite{SrSa2020, BSR2021}. They trained a \ac{NN}-based \ac{IDS} and proposed an architecture that addresses the requirements for the case of collision avoidance using vehicle-to-vehicle mechanisms. The main components of the architecture are: (1) a predictor of abnormal behavior that uses the trained \ac{IDS}, (2) machine learning models for computing the application decision trained from driving a vehicle in different weather conditions (windy, raining, snowing, and clear) and road types (city, suburban, and highway) which are used to estimate the response of the module, and (3) a plausibility check module that checks the safety of using the output of the response estimator. When the system detects an anomaly, it replaces the response computed by the collision application with the output of the response estimator when the plausibility check is positive; that is, the estimated response is safe. The system triggers service degradation if it detects an anomaly and the plausibility checks of the output of the estimated response is negative.

\section{Intrusion Detection System}\label{sec:IDS}


\acp{ECU} collaborate to perform tasks such as increasing speed, break, etc., by sending messages through the \ac{CAN} bus~\cite{9076852}. Attackers take control of a connected vehicle by injecting messages into its CAN bus. To mitigate such attacks, we have developed an \ac{ML}-based \ac{IDS} that captures the pattern of the sequences of the \ac{CAN} messages and represent them with a directed graph, which we call \emph{\acf{MSG}}, where the nodes represent the CAN IDs of the messages and the edges represent the sequences order of the messages, as depicted by Figure~\ref{fig:messagesgraph}~\cite{9490207}.

To construct the \ac{MSG}, we first create a dictionary of the CAN IDs exchanged in the \ac{CAN} bus. Then, we label the nodes of the \ac{MSG} with the CAN IDs and the edges with value "0". Next, we loop over the batch of the CAN messages of size $w$ (e.g., 1000 successive messages) that were exchanged from time $t$. For each of the messages, we increase the label of the edge linking the node representing the CAN ID of the message to the node repressing the CAN ID of the previously processed message. Equation~\ref{eq:disribution} represents the distribution of the messages-sequences at time $t$.

\begin{equation}\label{eq:disribution}
    D(t) = \{E(N_i,N_j)(t) \}
\end{equation}

    where $N_k$ is for node $k$ and $E(a,b)$ is for the edge from node $a$ to node $b$. \\
    
We consider that a \ac{MSG} representing the $w$ messages exchanged in a CAN bus at time $t$ is $similar$ to the \ac{MSG} representing the $w$ CAN messages exchanged during the following time slot $t+1$ in the case of normal driving behavior and that injection of messages into the CAN bus disrupts this pattern~\cite{9490207}--see Figure~\ref{fig:messagesgraph}~\cite{9490207}. Equation~\ref{eq:Similarity} formulates the \textit{Similarity} concept for the \ac{IDS}. That is, the similarity \textit{Sim} at time \textit{t+1} is the similarity of the distributions of the messages-sequences $D$ at time \textit{t} and at time \textit{t+1}. 

\begin{equation}\label{eq:Similarity}
    Sim(t+1) = Similarity(D(t),D(t+1))
\end{equation}

We use cosine similarity to measure the similarity between two \acp{MSG} of two successive time steps $t$ and $t+1$. The metric measures the angle between two vectors, where the closer the value is to 1, the more similar the two vectors are. Then, we use \emph{similarity threshold} technique to identify CAN message injections. The technique provides an accuracy of 97.32\%/--see Ref.~\cite{9490207} for details about the construction and accuracy of the technique. 

\begin{figure}
  \includegraphics[width=.5\textwidth]{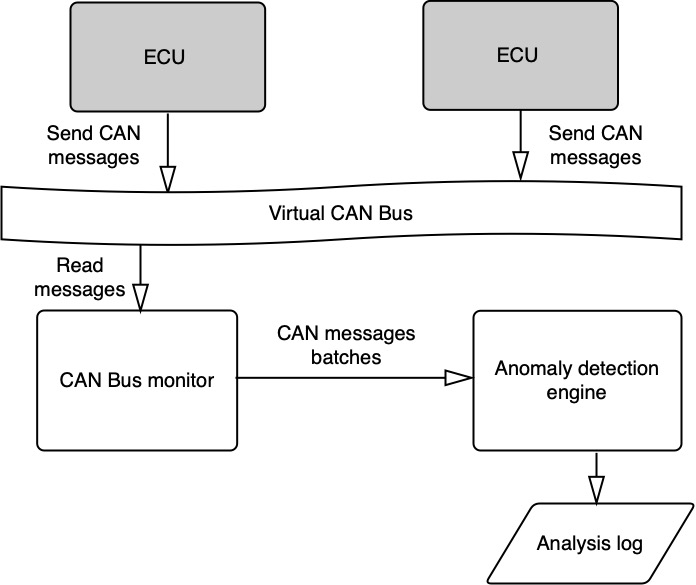}
  \caption{High-level architecture of the system.}
  \label{fig:architecture}
\end{figure}

We discuss in the following sections the architecture alternatives for monitoring the \ac{CAN} Bus and using the similarity threshold technique to identify  in real-time malicious injections of CAN messages.  

 \begin{table}[tbp]
 \begin{threeparttable}
\caption{IDS architecture constraints.}
\label{tab:ArchitectureConstraints}
\centering
\begin{tabular}{p{2.1in} p{1.0in}}
\rowcolor{Gray}\hline
  Constraint  & Value\\
\hline
Recommended maximum rate of injection of CAN messages  & 1908/sec \\\hline
Rate of injection of CAN messages  & 1000/sec~\cite{9076852} \\\hline
Reaction time constraint & 2.5 sec \\\hline
Detection speed of 1000 messages using the similarity threshold technique& 0.094 seconds\\\hline
\hline
\end{tabular}
    \begin{tablenotes}
      \item - The CAN bus is designated for a maximum signaling of 1 Mbits/s~\cite{TeIn2016} but 250kb/s is the recommended rate by the \ac{SAE} in J1939 standards~\cite{J1939}, which transports up to 1908 CAN frame per second -- 1908=250000/(128+3) where the data payload is of 8 bytes.
      \item - The break reaction time is less than 2.5 second for 90\% of the drivers.\cite{breakreact}
      \item - We postulate that the IDS needs to process the CAN messages batch file in less than 2.5 seconds, which is the upper bound of breaking reaction time~\cite{breakreact}.
    \end{tablenotes}
  \end{threeparttable}
\end{table}

\section{Real-time architecture alternatives of \ac{IDS} for Connected Vehicles}\label{sec:arechitecture}

In a typical IDS environment, a set of \acp{ECU} inject CAN messages into the CAN bus, a CAN Bus monitor captures the messages exchanged in the bus, and the anomaly detection engine analyzes these messages to identify malicious message injections, as depicted in Figure~\ref{fig:architecture}. The \emph{\ac{CAN} bus monitor} reads the messages available in the CAN bus continuously and sends them to the anomaly detection engine in batches of, e.g., 1000 messages. The \emph{anomaly detection engine} constructs a \ac{MSG} from the received batch of messages, applies the cosine similarity threshold attack detection technique, and outputs the results of the evaluation.

The \ac{IDS} must address three main requirements depicted in Table~\ref{tab:majorRequirements}. First, the response time of the \ac{IDS} must be less than the expected breaking reaction time; satisfying this requirement makes the \ac{IDS} a good candidate for e.g., a safety resilience mechanism that activates the breaks in case of detection of cyber-attacks. Second, the loss of CAN messages is not allowed, which is important for the reliability of IDS. Third, The IDS must run on a cheap ECU that has limited capabilities in terms of processor speed and memory size. In addition, Table~\ref{tab:ArchitectureConstraints} enumerates a set of constraints that the architecture shall satisfy. Since the goal of this paper is to assess architectures of real-time \ac{IDS}, we use the existing anomaly detection module developed in previous work~\cite{9490207} and do not create new ones.

 \begin{table}[tbp]
\caption{Practical concurrency scenarios of the IDS main components.}
\label{tab:ArchitectureScenarios}
\centering
\begin{tabular}{p{1.4in} p{.55in}p{.6in}p{.22in}}
\rowcolor{Gray}\hline
  Architecture scenario & Concurrency technique of CAN Bus monitor & Concurrency technique of the anomaly detection agent & Use of a queue\\
\hline
Scenario 1 - The two components run in a single process & main process  & main process  & no \\\hline
Scenario 2 -The two components run in a single process that includes one sub-process s& main process & child-process&no \\\hline
Scenario 3 - The two components run in a single process with two threads & thread & thread &yes\\\hline
Scenario 4 - The two components run in two processes & main process & main process & yes \\\hline
\hline
\end{tabular}
\end{table}

\begin{figure}
\centering
  \includegraphics[width=.45\textwidth]{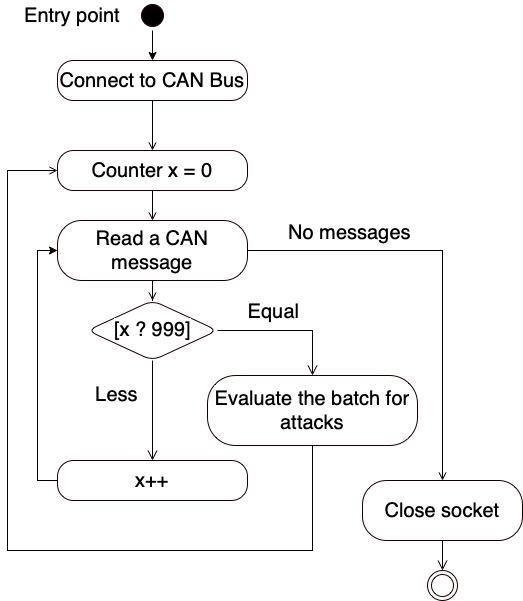}
  \caption{IDS flowchart diagram.}
  \label{fig:flowchart}
\end{figure}

\begin{figure}
\centering
  \includegraphics[width=.45\textwidth]{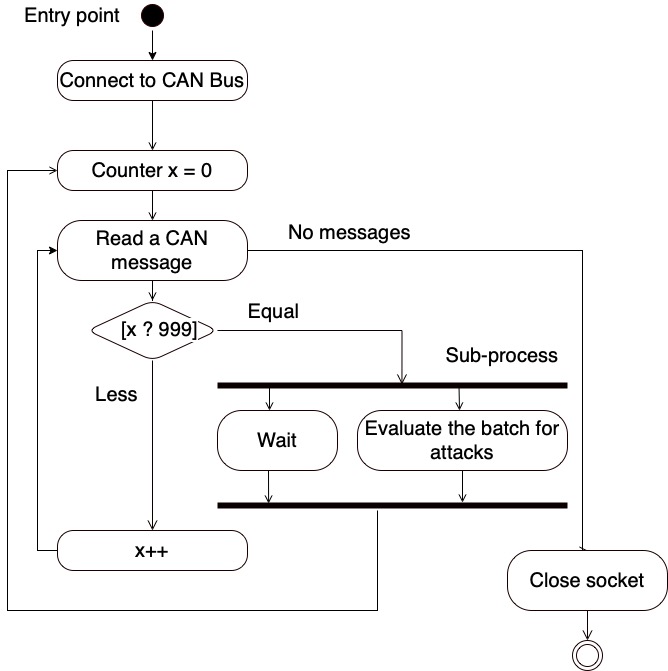}
  \caption{Flowchart diagram of the IDS architecture of scenario 2.}
  \label{fig:architectureScenario2}
\end{figure}

To satisfy the first requirement, we implement the \ac{CAN} Bus monitor and the anomaly detection engine using the C language. The anomaly detection engine uses PyObject library to call the data analysis module, passing it the CAN messages batches as a parameter, and parsing the analysis result. 

Typically, each \ac{ECU} services/reacts to each message it reads from the CAN bus (process or ignore) at a rate higher than the sending messages rate, to avoid loss of messages. In contrast, the IDS techniques process the CAN messages in batches, which takes much longueur than one millisecond. Thus, the \ac{IDS} would lose data if the CAN bus monitor and anomaly detection engine operate sequentially, as depicted in Figure~\ref{fig:flowchart}, which violates the third requirements that we set for our real-time \ac{IDS}: loss of CAN messages is not allowed. This problem could potentially be addressed by running the CAN Bus monitor and anomaly detection engine concurrently. The three concurrency techniques that could be used are: using separate processes, using sub-processes, and using threads. Table~\ref{tab:ArchitectureScenarios} shows the four potential architecture scenarios for using concurrency techniques of the CAN Bus monitor and the anomaly detection engine components to address the second and third requirements discussed above. We discuss in the following each of the architecture scenarios. 

\noindent {\bf Scenario 1 - The two components run in a single process.} In this scenario, the CAN Bus monitor and anomaly detection engine run sequentially as depicted by Figure~\ref{fig:flowchart}. The CAN Bus monitor can potentially lose CAN messages that the \acp{ECU} send while the IDS busy analyzing the CAN messages batch it receives to identify potential attacks.  

\noindent {\bf Scenario 2 - The two components run in a single process that includes one sub-process.} The \ac{CAN} bus monitor and the anomaly detection agent run in a single process. As depicted by Figure~\ref{fig:architectureScenario2}, the main process continuously reads the CAN messages from the CAN bus and creates a sub-process, that evaluates the batch for attacks, that is executed when there are enough messages for the batch. Note that the sub-processes become zombies at the end of their executions, and it is complicated to shut them down from the main processes.

\begin{figure}
\centering
  \includegraphics[width=0.45\textwidth]{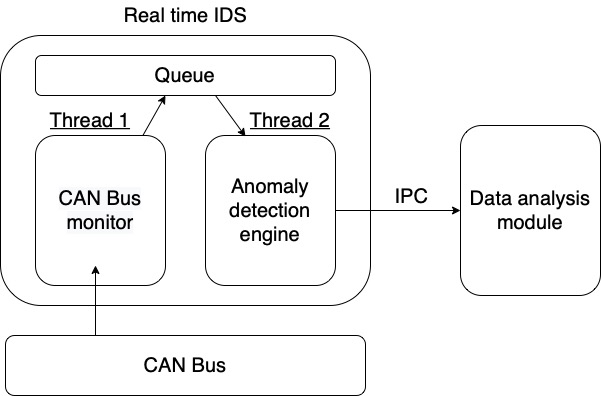}
  \caption{Block diagram of the IDS architecture of scenario 3.}
  \label{fig:architectureScenario3}
\end{figure}

\noindent {\bf Scenario 3 - The two components run in a single process with two threads.} The \ac{CAN} bus monitor and the anomaly detection agent run in a single process but in separate threads, as depicted by Figure~\ref{fig:architectureScenario3}. We use a queue to pass data between the two components to prevent losses of CAN messages. The anomaly detection engine uses \ac{IPC} technique to execute the data analysis module, which implements the technique discussed in Section~\ref{sec:IDS}.

For completeness, there are two other variants of using threads with a single process besides the one discussed above, which are: (1) implementing the CAN bus monitor in a thread and the anomaly detection engine in the main process and (2) implementing the anomaly detection engine in a thread and the CAN bus monitor in the main process. We do not report the evaluation of these variants because they use the same concurrency technique and do not outperform the two threads variant.

\begin{figure}
\centering
  \includegraphics[width=0.45\textwidth]{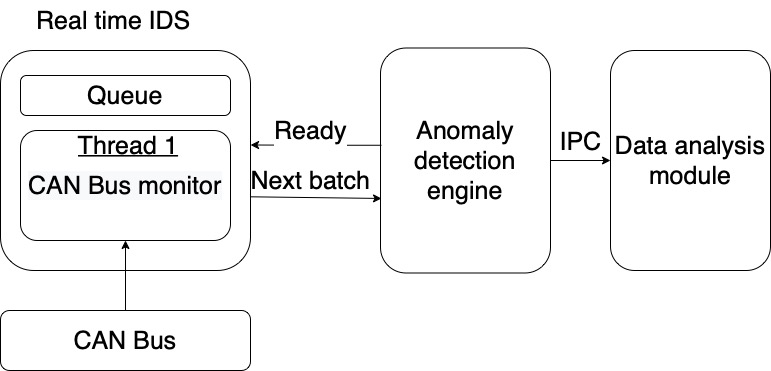}
  \caption{Block diagram of the IDS architecture of scenario 4.}
  \label{fig:architectureScenario4}
\end{figure}

\noindent {\bf Scenario 4 - The two components run in two processes.} The \ac{CAN} bus monitor and the anomaly detection agent run in separate independent processes, as depicted by Figure~\ref{fig:architectureScenario4}. We also use a queue to pass data between the
two components to prevent losses of CAN messages.

\section{Evaluation of the four real-time IDS architecture scenarios}\label{sec:evaluation}

This section describes the evaluation environment and datasets, compares the anomaly evaluation times of the four architecture scenarios, and analyses the impact of CAN message injection speed on the message loss ratio of the four architecture scenarios.   

\subsection{Evaluation environment and datasets}

\begin{figure}
  \includegraphics[width=.4\textwidth]{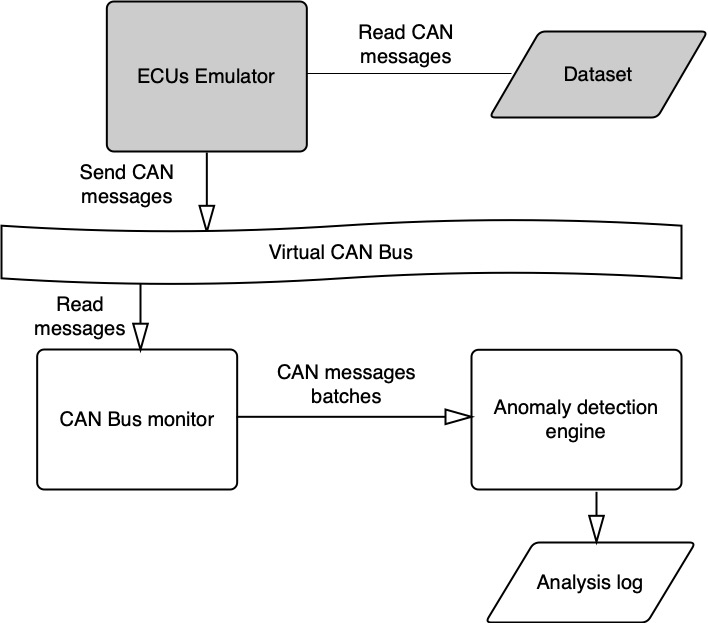}
  \caption{Block diagram of the evaluation environment of the real-time IDS architecture scenarios.}
  \label{fig:evaluationarchitecture}
\end{figure}

We implemented the four architecture scenarios and deployed them to a Raspberry Pi that runs Raspbian 10, with four core processors of 1.2 MHz speed and 1GB of memory. Figure~\ref{fig:evaluationarchitecture} shows the evaluation environment of the architecture scenarios. The environment uses an \emph{ECUs emulator} that simulates car \acp{ECU}, which sends messages periodically into Linux virtual CAN bus. 

\begin{table}[tbp]
\caption{CAN messages datasets}
\label{tab:listdatasets}
\centering
\begin{tabular}{p{0.1in} p{2.2in}p{.6in}}
\hline
\rowcolor{Gray}\hline
No &  Description & \# of \ac{CAN} messages \\\hline
\hline
  1   &  CAN Data for no injection of fabricated messages  & 23,963 \\\hline
  \rowcolor{lightgray}
  2   &  CAN Data with injection of "FFF" as the speed reading&  88,492 \\\hline
 3& CAN Data with injection of "FFFF" as the RPM reading & 30,308\\\hline
\hline
\end{tabular}
\end{table}

\begin{table}[tbp]
\caption{Speed of simulating injection of CAN messages and evaluating them for attacks using architecture scenario 1.}
\label{tab:injectionanddetctionspeed}
\centering
\begin{tabular}{p{0.7in} p{0.7in}p{0.7in}p{0.6in}}
\rowcolor{Gray}\hline
    & Normal messages & messages with injection of seed readings & messages with injection of RPM readings\\
\hline
\multicolumn{4}{l}{Time to send 1000 CAN messages in seconds	}\\	\hline	
	
Min	& 0.946	&	0.925	&	1.015 \\\hline
Max	& 1.120	&	1.217	&	1.056 \\\hline
Average	& 0.992	&	1.008	&	1.023 \\\hline\hline
			
\multicolumn{4}{l}{Time to evaluate 1000 CAN messages in seconds} \\\hline		\hline

Min	&0.115		&0.116	&	0.131\\\hline
Max		&0.271	&	0.274	&	0.180\\\hline
Average		&0.154	&	0.143	&	0.151\\\hline
\hline
\end{tabular}
\end{table}

In this evaluation, we use datasets~\cite{othmane2020b} of CAN bus messages for (1) normal driving behavior, (2) injection of fabricated speed reading messages onto the CAN bus, and (3) injection of fabricated RPM reading messages onto the CAN bus collected from an in-motion Ford Transit 500 2017 ~\cite{9076852}. Table \ref{tab:listdatasets} lists the datasets and the number of CAN messages in each of them. 

The \emph{ECUs emulator} reads the CAN messages stored in the dataset files and sends them through the virtual CAN bus. The messages are processed by the CAN bus monitor and anomaly detection engine in the four architecture scenarios. Table~\ref{tab:injectionanddetctionspeed} provides the time that the ECUs emulator takes to send 1000 messages into the CAN Bus and the time that the IDS takes to evaluate 1000 CAN messages for the cases of normal messages dataset, messages with the injection of speed readings dataset, and messages with the injection of RPM readings dataset. We do not observe a big difference in processing a batch of normal CAN messages, messages with the injection of speed readings, and messages with the injection of RPM readings. We observe that the IDS takes an average 149 milliseconds to evaluate a batch of 1000 messages, while the time to send 1000 messages into the CAN bus is about 1,007 seconds. The IDS would lose about 149 CAN messages from each batch, which can impact the attack detection rate--i.e., ignoring 14,9\% of the messages can impact the detection rate.   

\begin{figure}[tb]
        \centering
        \includegraphics[width=1.0\linewidth]{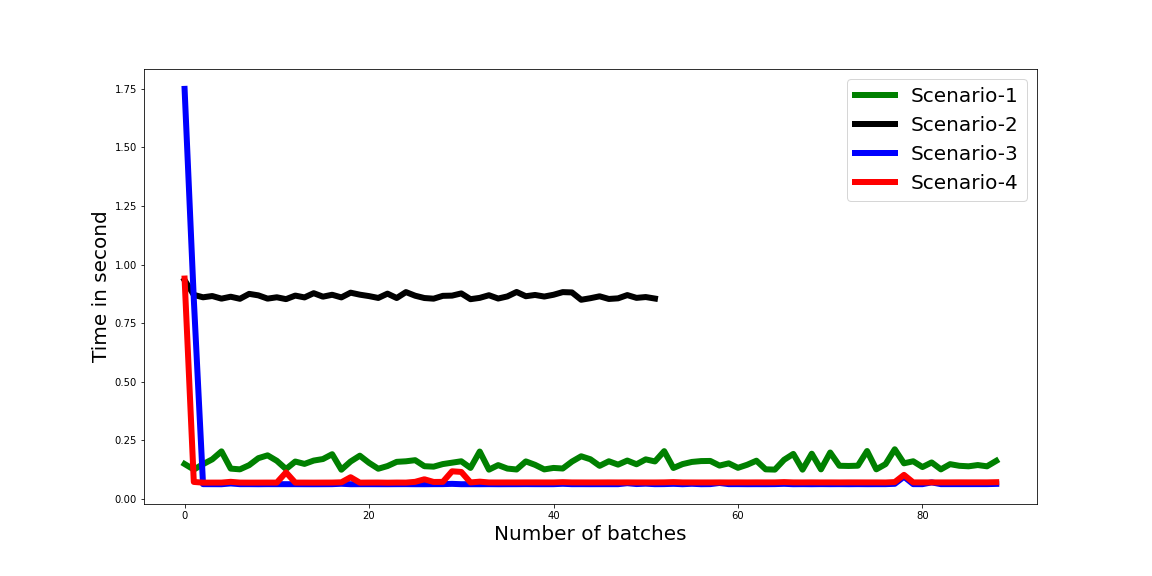}
    \caption{Anomaly evaluation time of the four architecture scenarios.}
    \label{Fig:responsetime}
\end{figure}

\begin{table}[tbp]
\caption{Anomaly evaluation times and response times for the four architecture scenarios.}
\label{tab:responsetime}
\centering
\begin{tabular}{p{1.3in} p{.6in}p{.4in}p{.4in}}
\rowcolor{Gray}\hline
  Architecture & Average time of sending 1000 CAN messages & Average evaluation time& Response time\\
\hline
Scenario 1 - Single process with no threads &998 ms & 152 ms& 1.15 sec.  \\\hline
Scenario 2 - Single process with one sub-process& 944 ms & 865 ms& 1.809 sec \\\hline
Scenario 3 - Single process with two threads & 950 ms & 90 ms& 1.04 sec.\\\hline
Scenario 4 - Two processes &945 ms  &81 ms& 1.026 sec.\\\hline\hline

\end{tabular}
\end{table}

\subsection{Analysis of the anomaly evaluation time of the four architecture scenarios}

We configured the ECU emulator to send the dataset "CAN Data with injection of "FFF" as the speed reading" with a rate of about 1000 messages through the virtual CAN Bus. Table~\ref{tab:responsetime} shows the average time of sending a batch of CAN messages, the average evaluation time, and the average response time (time taken from collecting the first CAN message of the batch to output the result of the evaluation of the batch) of the four IDS architecture scenarios. We consider the \emph{anomaly evaluation time} as the difference between the end of evaluating the messages batch for attacks and the end of reading 1000 CAN messages from the virtual CAN Bus. The data shows that anomaly evaluation time is way below the batch messages sending time. Figure~\ref{Fig:responsetime} shows the anomaly evaluation time of the four selected real-time \ac{IDS} architecture scenarios. The diagram shows that anomaly evaluation times of a real-time \ac{IDS} designed as a single process, single process with two threads, and as two processes are below the brake reaction time (0.5 to 2.5 seconds), In addition, the anomaly evaluation times of the \ac{IDS} designed as a single process with two threads and as two processes are too close. 

We conclude that the IDS response time of a real-time \ac{IDS} designed as a single process with no threads, as a single process with two threads, and as two processes is below the brake reaction time (assuming the rate of sending messages through the CAN Bus is about 1000 messages/second) which makes them a good candidate for real-time IDS for connected vehicles.    

\begin{figure}[tb]
        \centering
        \includegraphics[width=1.0\linewidth]{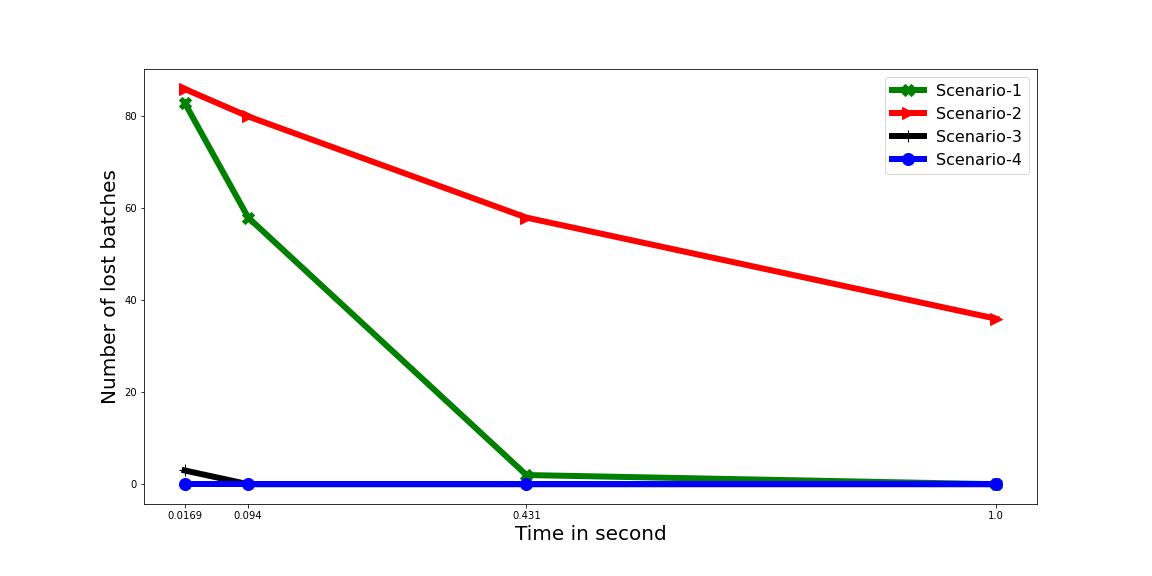}
    \caption{Ratio of messages losses vs speed of sending 1000 CAN messages through the CAN bus in the four architecture scenarios.}
    \label{Fig:messagelossratio}
\end{figure}

\subsection{Analysis of the reliability of the four architecture scenarios in terms of CAN message losses}

 \emph{Message loss ratio} is the ratio of CAN messages that are sent through the CAN Bus by the ECUs emulator but are not received and processed by the anomaly detection engine. Theoretically, the CAN messages that the ECUs emulator sends through the CAN Bus while the single process IDS architecture is busy analyzing a batch of previously received messages are lost. Figure~\ref{Fig:messagelossratio} shows the relationship between the messages loss ratio and speed of sending messages into the virtual CAN bus by the ECUs emulator. The figure shows that the ratio of messages loss decreases as the speed of sending CAN messages increases and it reaches 0\% for the case of architecture scenario 1 and 2 and that there are no messages losses for the case of \ac{IDS} designed as as two processes--because the servicing time is smaller than the time to send the messages batch into the CAN Bus.  

\section{Conclusion}\label{sec:conclusion}

Machine-learning-based \acp{IDS} techniques have been proposed for the detection of malicious injection of messages into the in-vehicle network of connected vehicles. Evaluations of such \ac{IDS} have been performed offline, which limits the confidence in their use for real-time protection scenarios. We evaluated in this paper four architecture designs for real-time \ac{IDS} for connected vehicles using an anomaly detection engine that uses similarities of graphs representing sequencing of CAN messages during a given period and a \ac{CAN} dataset collected from a moving vehicle under malicious speed reading message injections. The evaluation shows that a real-time \ac{IDS} for a connected vehicle designed as two processes are reliable (no loss of messages) and have a low anomaly evaluation time that makes them a good candidate for real-time resilience mechanisms as a response to cyber-attacks.

\section{Acknowledgement}

The authors thank Arun Somani from Iowa State University for the thorough discussions about the research. This research is partly funded by Iowa State University's Regents Innovation Fund (RIF).

\bibliographystyle{ieeetr}
\bibliography{RealTimeIDS} 

\end{document}